\documentclass[twocolumn,showpacs,amsmath,amssymb,prl,superscriptaddress]{revtex4}
\usepackage{graphicx}
\usepackage{bm}

\begin{document}
\title{Phase diagram for Coulomb-frustrated phase separation in systems
  with negative short-range compressibility}

\author{C.  Ortix}
\affiliation{Dipartimento di Fisica, Universit\`{a} del Salento and INFN Sezione di Lecce, Via per
  Arnesano, 73100 Lecce, Italy. }
\author{J.  Lorenzana} 
\affiliation{SMC-INFM and Dipartimento di Fisica, Universit\`a di Roma 
``La Sapienza'', P.  Aldo Moro 2, 00185 Roma, Italy.}
\affiliation{ISC-CNR, Via dei Taurini 19, 00185 Roma, Italy.} 
\author{C.  Di Castro} 
\affiliation{SMC-INFM and Dipartimento di Fisica, Universit\`a di Roma 
``La Sapienza'', P.  Aldo Moro 2, 00185 Roma, Italy.}
\date{\today}

\begin{abstract}
Using numerical techniques and asymptotic expansions
we obtain the phase diagram of a paradigmatic model
of Coulomb frustrated phase separation in systems with negative short-range compressibility. 
The transition from the homogeneous phase to the inhomogeneous phase
is generically 
first order in isotropic three-dimensional systems except for a
critical point. Close to the critical point, inhomogeneities are
predicted to form a BCC lattice with subsequent transitions to a
triangular lattice of rods and a layered structure.  
Inclusion of a strong anisotropy allows for second- and
first-order transition lines joined by a tricritical
point.
\end{abstract}
\maketitle
The appearance of spatial inhomogeneities is a general phenomenon
occurring in a wide variety of systems with competing interactions on
different length
scales\cite{seu95,mur02,don95,cas95b,mar90,kug05,fis07,nag67nag83,fre02,kag99,oht86,low94,sch00,nus99,tar06,sci04,ort07b}.
In electronic systems strong electron 
correlations often drive the electronic compressibility
negative as in the Hubbard model in infinite dimension\cite{don95}, the
Falicov Kimball model\cite{fre02}, the electron
gas\cite{mah00,neutrons}, cuprates models\cite{cas95b,mar90,eme90c}, 
manganite models\cite{nag67nag83,kag99,kug05},
semiconductor heterostructures\cite{fis07}, to mention a few\cite{mur02}.
This signals a tendency to phase separation at macroscopic lengths
which is frustrated by the long-range Coulomb interaction.  
Although there is agreement that the system can become inhomogeneous
at a mesoscopic scale
if the Coulomb frustration is not too large, the nature of the
transition has not yet been settled. This is an important question
in general but 
particularly so in metallic systems where the existence of a second
order quantum critical point
separating an homogeneous phase from an inhomogeneous one can disrupt 
the Fermi liquid behavior\cite{cas95b}.

\begin{figure}[tbp]
\includegraphics[width=7 cm]{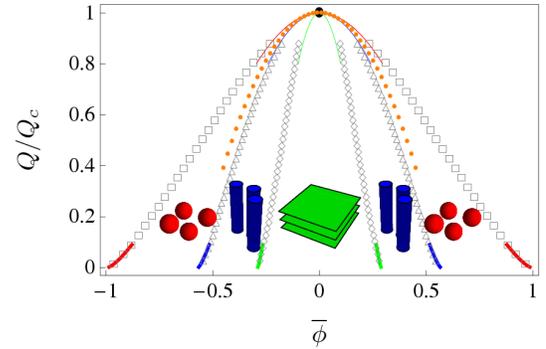}
\vskip -0.3cm
\caption{(Color online) Phase diagram in three-dimensional isotropic
  systems.  The small dots indicate the Gaussian instability line $Q_g$.
  The thin (thicks) lines represent first-order transitions
  in the strong (weak) coupling approximation. In the two limits they
  overlap with the 
  corresponding numerically determined transition lines  
  from homogeneous to droplet-like 
  inhomogeneities ($\square$),  from droplets to rods ($\triangle$),
  and from rods to  layers ($\diamondsuit$).}  
\label{fig:3dphasediagram}
\end{figure}

\begin{figure}
\begin{center}
\vspace{-.5cm}
\includegraphics[width=7 cm]{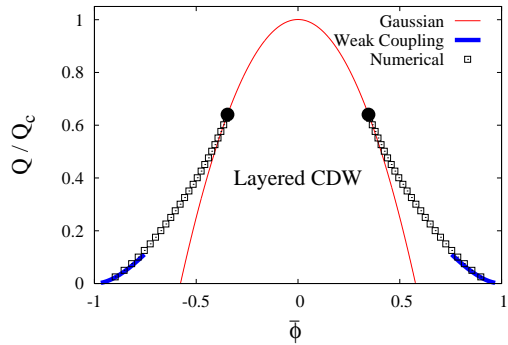}
\vspace{-.5cm}
\end{center}
\caption{(Color online) The phase diagram for anisotropic 3D systems. The thin line
  corresponding to the Gaussian instability determines the second
  order transition line above the tricritical point ($\bullet$). Below
  the tricritical point the transition is first order and is
  determined  numerically ($\square$) and in the weak coupling limit assuming
  sharp interfaces\cite{mur02,ort08} (thick line). Below the
  tricritical point the  Gaussian line becomes the limit of
  metastability of the uniform   phase.  } 
\label{fig:phasediagramexact}
\end{figure}

 In this work we study a  $\phi^4$ model augmented with long-range
 interactions as a generic model of Coulomb frustrated phase
 separation in systems with negative short-range compressibility. 
The model (or closely related variants) has been used to describe
 inhomogeneities in a variety of 
 systems\cite{ort07b,oht86,low94,sch00,nus99,tar06} including 
 mixtures of block copolymers \cite{oht86}, 
charged colloids in polymeric solutions \cite{tar06} and electronic
 systems\cite{low94,sch00}. 
We determine the transition line from the homogeneous state to the
inhomogeneous state using numerical and analytical techniques and show
that generically the transition is first order in three spatial 
dimensions (3D) except for a critical point
(Fig.~\ref{fig:3dphasediagram}). This outcome changes if a strong anisotropy
is taken into account. Then both first and second order transitions are
allowed separated by a tricritical point
(Fig.~\ref{fig:phasediagramexact}). In addition we study the crossover
from harmonic to anharmonic inhomogeneities and the different
topological transitions of the 3D isotropic system. Close to the critical
point the inhomogeneities are shown to form a BCC lattice.

The model is defined by  the following Hamiltonian 
\begin{eqnarray}
{\cal H} & =& \int d  {\bf x} \left[\phi({\bf x})^{2}-1\right]^2+|\nabla \phi({\bf x})|^{2}\label{eq:eGLrescale} \\
& +  & \frac{Q^2}{2 } 
\int d {\bf x}\int d {\bf x^{\prime}} \dfrac{\left[\phi \left({\bf x}\right)-
  \overline{\phi}\right] \left[\phi \left({\bf x^{\prime}}\right)- \overline{\phi}\right]}{\left|{\bf x-x^{\prime}}\right|},  \nonumber
\end{eqnarray}
with the scalar classical field $\phi$ representing the local charge
density, $\overline \phi$ the average density. A rigid background
ensures charge neutrality.   
In the case of electronic systems with a negative compressibility the
model describes phenomena at large length scales compared to the
underlying lattice constant. It can be derived by expanding the coarse
grained energy of the system 
around a reference density belonging to the negative compressibility density
region. Fermionic effects like the Pauli exclusion principle are
reflected in the coefficients.
The reference density can be fine tuned so as to eliminate 
a small cubic term. (Our results can be trivially extended to take
into account a large cubic term).   Constant and linear
terms are inessential,  thus one obtains a double well
energy which can be taken to the dimensionless form of
Eq.~\eqref{eq:eGLrescale} by measuring energies
in terms of the barrier height, distances in terms of the bare
correlation length $\xi$, and density $\phi$ in units such that the double
well minima are at $\phi=\pm1$. This leads to a renormalized
Coulomb coupling in 3D, $ Q^{2}\equiv 4 e^{2} \xi^{2}/(\epsilon_{0}
|a|)$. Here $e$ is the electron charge, 
$\epsilon_0$ is a dielectric constant due to external
degrees of freedom and $a<0$ is proportional to the inverse
short-range compressibility. More precisely,  it is  the second
derivative of the short range part of the energy per unit volume  
with respect to the density at the reference density.

In general $a$ and $\xi$ will depend on external parameters like
pressure. They can even be taken as temperature dependent, as in
Landau theory, in which case 
${\cal H}$ has to be interpreted as a free energy and 
the model becomes
a mean-field description of a temperature driven transition to an
inhomogeneous state. This can be useful, for example, to model
inhomogeneities appearing below some temperature in manganites\cite{zha02}.

\begin{figure}
\includegraphics[width=9 cm]{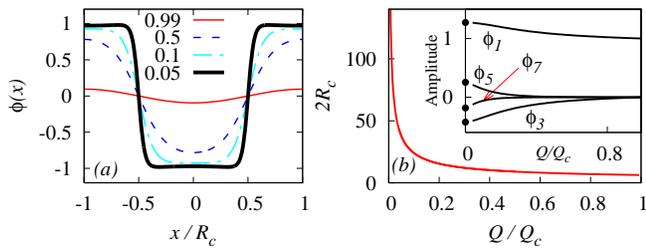}
\vspace{-.8cm}
\caption{(Color online) 
(a) Behavior of the charge density modulation for $\overline \phi=0$
 and different couplings (labeled by the value of $Q/Q_c$).
 Near the Gaussian instability the modulation
is close to a  SCDW (thin full line). As $Q$ decreases unharmonicity
 is build in and the charge modulation tends to  a square wave which
 has as limiting case Maxwell construction at $Q=0$. 
 (b) Evolution of the periodicity of the charge density wave measured in
units of $\xi$. 
Inset:
Evolution of the harmonic amplitudes $\phi_{n} / \sum_{n^{\prime}>0}
 \phi_{n^{\prime}}$ as a function of $Q$. The dots 
 indicate the amplitudes for a layered structure of $\phi=\pm 1$ as
 ruled by Maxwell construction at $Q=0$. Amplitudes for even $n$ vanish by
 symmetry.  }  
\label{fig:chargeden}
\end{figure}

By computing the static response to an external field in momentum
space, we get the charge susceptibility at finite momentum ${\bf k}$
which depends only on the modulus of the momentum, $k$, measured in
units of $\xi^{-1}$:  
\begin{equation}
\chi ({\bf k})= \left[ k^{2}+\frac{2 \pi Q^{2}}{ k^{2}}-2+6 \overline{\phi}^{2}\right]^{-1}\;\;\;\;\;({\bf k}\ne0)
\label{eq:chidef}
\end{equation}
The susceptibility has a maximum at 
$k_{0}=\left[2 \pi Q^{2}\right]^{1/4}$ which 
diverges as $Q$ approaches the Gaussian instability line $Q_{g}$ (the dotted line in Fig.~\ref{fig:3dphasediagram}) from
above, where   $Q_{g}=Q_c\left[1-3\overline \phi^{2}\right]$ and
$Q_c=1/\sqrt{2\pi}$. 
 This indicates an  instability of the homogeneous phase toward a
 sinusoidal charge density wave  (SCDW) of periodicity
 $2R_{c}=2\pi/k_{0}$
  with vanishing wave amplitude at the transition and direction chosen
 by spontaneous symmetry breaking. 

The Gaussian transition can not survive at 
low $Q$. Indeed as $Q\rightarrow 0$ the Gaussian theory predicts
 inhomogeneities in the range 
$|\overline\phi|<1/\sqrt{3}$
as opposed to Maxwell construction at $Q=0$ which predicts a globally  
inhomogeneous state in the range  $|\overline\phi|<1$. We now show
that the system never 
 reaches the Gaussian instability except for a critical point in the phase
 diagram: the Gaussian line is preempted by a first order transition.
Restricting to periodic textures, the free energy difference respect to
the uniform state can be written in Fourier space as: 
\begin{eqnarray}
\delta F&=&\dfrac{1}{V}\sum_{{\bf G} \neq 0} \phi_{{\bf G}}
\chi^{-1}\left({\bf G}\right)\phi_{-{\bf G}} \nonumber \\ 
& +& \dfrac{4 \overline{\phi}}{V^{2}}\sum_{{\bf G}_{1},{\bf G}_{2},
  {\bf G}_{3}\neq 0} \,\phi_{{\bf G}_{1}}\phi_{{\bf G}_{2}}\phi_{{\bf
    G}_{3}} \delta_{{{\bf G}_{1}+{\bf G}_{2}+{\bf G}_{3}, 0}}
\label{eq:fGLfinal}  \\ 
&+&\dfrac{1}{V^{3}} \sum_{{\bf G}_{1},{\bf G}_{2},{\bf G}_{3},{\bf G}_{4}\neq 0} \phi_{{\bf G}_{1}} \phi_{{\bf G}_{2}} \phi_{{\bf G}_{3}} \phi_{{\bf G}_{4}}  \delta_{ {{\bf G}_{1}+{\bf G}_{2}+{\bf G}_{3}+{\bf G}_{4}, 0}} \nonumber
\end{eqnarray}
where the ${\bf G}$'s are the reciprocal lattice vectors and $V$ is
the unit cell volume of the periodic solution.

The appearance of a self-generated cubic term in Eq.~\eqref{eq:fGLfinal} calls for the
possibility of a first-order transition which can be treated
analogously to the liquid-solid transition\cite{ale78,cha95}. 
Assuming that the instability is weakly first order, which can be
checked a posteriori, we restrict the sum to  
wave-vectors with magnitude $G$ close to $k_0$.   
To have an energetic advantage from the cubic term of
Eq.~(\ref{eq:fGLfinal}) we need to find reciprocal lattices with  
triads of wave-vectors forming equilateral
triangles so that they add to zero. 
 By requiring symmetric
structures with inversion only three sets of vectors  $L_G$
are allowed that correspond to FCC, planar hexagonal, and icosahedral
reciprocal lattices\cite{cha95}. 
The free energy reads:
\begin{equation}
  \label{eq:felatt}
\dfrac{\delta F}{V}= \chi^{-1} \left(G\right) m \phi_{G}^{2}+ 8 \overline{\phi} \,p \, m \phi_{G}^{3}+\phi_{G}^{4} \left[3 m\left(m-1\right)+6 q m\right]  
\end{equation}
where the  Fourier component amplitudes $\phi_{G}$ depends only on the
 modulus of ${\bf G}$,
 $m$ is the number of vectors in $L_G$ and $p$ ($q$) is the
number of triangles (non-planar diamonds) to which each vector
belongs when the triangles are accommodated in regular geometrical objects 
\cite{note}.
Upon minimizing Eq.~\eqref{eq:felatt} with respect to the
wave-amplitude $\phi_{G}$ and $G$ one finds $G\equiv k_0$. Equating
the energies for the different structures one finds three first order  
transition lines between phases $X$ and $Y$
 of the form
\begin{equation}
Q_{X,Y}=Q_c (1- \alpha_{X,Y} \overline{\phi}^{2}).
\label{eq:transI3D}
\end{equation}
where $X,Y=U,B,T,L$ stand for uniform, BCC, triangular and layered
structures respectively (see below). The three first order transition lines 
[thin lines in Fig.~\ref{fig:3dphasediagram}] 
join at the critical point $(\overline\phi,Q)=(0,Q_c)$ shown with a
$\bullet$. At the critical point  
one recovers a SCDW second-order phase transition with a charge
susceptibility divergence since the cubic term of
Eq.~\eqref{eq:fGLfinal} vanishes. Away from the critical point the 
 Gaussian line is the limit of metastability of the
 homogeneous phase (doted line).  For all the phases, which we describe
 next, the order parameter at the transition goes linearly to zero as
 $\bar \phi \rightarrow 0$ confirming that the transition is 
 weakly first order close to the critical point.

Approaching the first order lines from above the 
first inhomogeneous structure to become
stable corresponds to the FCC reciprocal lattice defined by the $m=12$
wave-vectors  
${\bf G}\sqrt{2}/G  = \left(\pm 1,\pm 1,0\right)$, $\left(\pm 1,0,\pm
    1\right)$, $\left(0,\pm 1, \pm 1 \right) $ for which
$p=q=2$.  This corresponds to a BCC crystal of inhomogeneities in real
space with $\alpha_{U,B}=103/45\approx 2.29$.

Decreasing $Q$ at finite $\overline \phi$  the planar hexagonal
lattice, with $m=6$, $p=1$, $q=0$ and $\alpha_{BT}\approx 3.44$, 
becomes stable corresponding to 
rod-like inhomogeneities forming a triangular lattice
similar to an Abrikosov lattice in a type II superconductor
(middle thin line in  Fig.~\ref{fig:3dphasediagram}). 
For weaker Coulomb couplings and close to $\overline \phi=0$  we find
a subsequent morphological transition (lower thin line) 
that restores the translational
symmetry in an additional  direction and leads to a layered 
structure ($m=2 $, $p=q=0$) with $\alpha_{T,L}=87/(19-6
\sqrt{6})\approx 20.22$.  We also find that the icosahedral
reciprocal lattice which corresponds to an icosahedral quasicrystal 
never becomes favorable.

The above transitions  [c.f. Eq.~\eqref{eq:transI3D}] are expected to
be asymptotically exact close to 
the critical point. Instead for  $Q\rightarrow 0$ 
they are clearly inaccurate since, still, the  range of stability of the
inhomogeneous state is smaller than Maxwell construction.
 In this limit the weak first-order character of the
transition is lost and more harmonics should be taken into account.

A good approximation for small $Q$ consists in assuming  
domains of uniform density of one or the other phase separated by 
sharp interfaces\cite{nus99,lor01I,ort07b,ort07,mur02}.  This is a reasonable approximation for
$Q \rightarrow 0$ since there is a strong separation 
($l_{S}>>l_{d}>>\xi$) between the 
typical interface scale length $\xi$, the typical size of the domains
$l_d\equiv \xi/Q^{2/3}$ 
and the screening length
$l_{S}\equiv \xi /( \pi^{1/2} Q)$ 
which controls the relaxation of the charge inside a domain. 
The transition lines in this approximation  are shown
with the thick lines at the bottom of
Fig.~\ref{fig:3dphasediagram}.   Again one finds the same topological
transitions as in strong coupling but now the inhomogeneities form
sharply defined spherical drops, cylinders  and layers. The lattice is
treated in the  Wigner-Seitz approximation (WSA)\cite{nag67nag83,mah00}.

In order to study the crossover from the weak to strong coupling
we have numerically minimized a discretized  version of the
energy, also in the WSA. For
rod-like and droplet-like inhomogeneities
we  assume respectively cylindrical and spherical symmetry in order to 
reduce the minimization procedure to a one-dimensional effective
problem. The first-order transition points numerically
located are shown with the $\square$, $\triangle$, $\diamondsuit$ in
Fig.~\ref{fig:3dphasediagram}. 
We find that as the coupling is decreased the size of the domains 
become much smaller than their distance which should make the
approximation particularly accurate at weak coupling. Despite of that,  
the numerical result converges to the asymptotic expansions
both at weak and strong coupling indicating a 
range of applicability of the WSA wider than expected. 

If one relaxes the WSA, the BCC lattice, found at strong coupling, is
expected to evolve into a BCC Wigner crystal of drops at weak coupling
which is the lowest energy Wigner crystal lattice\cite{mah00}. Other
lattices however (FCC, HCP) are very close in energy which suggests that
amorphous configurations will be very competitive as well. 

It is interesting to see how unharmonicity  is build into the system
as the coupling is decreased. In Fig.~\ref{fig:chargeden} (a) we show 
the charge profile for $\overline \phi=0$ and different couplings. 
The SCDW smoothly evolves into the domain morphology  that has as a
limiting case the 
macroscopically phase separated state at $Q = 0$. Notice that the
horizontal axis is 
normalized by the cell periodicity $2 R_c$, which becomes of the order of the linear size of
the system as $Q  \rightarrow 0$ [Fig.~\ref{fig:chargeden} (b)]. 

The inset of Fig.~\ref{fig:chargeden} (b) shows the behavior of
 the Fourier components.  Close to the second-order phase transition the 
order $n$ harmonics, $\phi_n\equiv\phi_{G_n}$ with $G_n=n \pi/R_c$, behave as $\phi_{n} \propto
(Q_{g}-Q )^{|n/2|}$. This follows from the fact that 
the modulation $\phi_n$ couples with  $(\phi_1)^{|n|}$ in the quartic term
of Eq.~\eqref{eq:fGLfinal}.  
 Higher harmonics proliferate as $Q $ is
decreased and converge to a rectangular profile corresponding to
macroscopic phase separation at $Q = 0$ (shown by $\bullet$ in the inset).

The phase diagram changes  dramatically if the gradient term is made
anisotropic i.e. $\xi_\parallel /\xi_\perp>1$ where  $\xi_\parallel$ ($\xi_\perp$) is
the bare correlation length in the ``hard'' (``soft'') direction.  
This can originate from an underlying crystal which
favors certain orientations of the interfaces. In
Fig.~\ref{fig:phasediagramexact} we show the 3D phase diagram 
with two hard directions  and one soft direction for
$\xi_\parallel/\xi_\perp\rightarrow \infty$ so that only one-dimensional
 modulations are allowed.  The units are fixed as before with 
 $\xi_\perp$ replacing $\xi$.

In this case the cubic term in Eq.~\eqref{eq:fGLfinal} has no effect
and one recovers the Gaussian instability where SCDW appears at a
second order transition in an extended range of coupling.  Thus, 
for a metallic system, Landau damping becomes relevant and one
should take the coupling to the fermions explicitly into account\cite{cas95b}.

For the same argument as
before the Gaussian line can not persist up to $Q=0$.  
We find, in fact a tricritical point at  
$\left(\overline \phi,Q \right)=\left(\pm \sqrt{3}/5,16/25
  Q_c\right)$ where the transition becomes first-order ($\bullet$ in
Fig.~\ref{fig:phasediagramexact}). 
 The position of the tricritical points and the behavior of the
charge density modulation around it, can be studied retaining only
two-harmonics in the order parameter with collinear wave-vectors : 
$$\phi\left(x\right)= \overline{\phi}+ 2 [\phi_{1} \cos{(G_1 x)}+\phi_{2} \cos{(G_2 x)}]$$
and assuming $\phi_{2}<<\phi_{1}$. 

By expanding the energy around the tricritical point and  minimizing
with respect to the second-harmonic amplitude $\phi_2$ one obtains an
effective Landau free energy expansion for the modulated phase in
terms of $\phi_{1}$ only,
\begin{equation}
\dfrac{\delta F_{II}}{V}= r \phi_{1}^{2}+ u_{4} \phi_{1}^{4}+ u_{6} \phi_{1}^{6}
\label{eq:landautricri}
 \end{equation}
with $r\equiv 4 \left(Q-Q_{g}\right)/Q_c$, $u_4\equiv 6-32
 \overline{\phi}^{2} Q_c/Q>0$ and $u_6$ a positive constant. This free
 energy has the canonical form for a tricritical point which as usual  
is determined by the vanishing of the quartic term on the Gaussian line\cite{cha95}.

 In the first-order transition region 
 ($u_{4}<0$), the appearance of the inhomogeneities  is determined by the
 coefficient $u_{6}$ in the free energy expansion,
 Eq.~\eqref{eq:landautricri}, whose precise value depends upon
 higher-order harmonic contributions. As before we have determined the
 first-order transition line by numerical minimization ($\square$ in
 Fig.~\ref{fig:phasediagramexact}) and from the 
weak coupling expansion assuming sharp interfaces\cite{mur02,ort08}
 (dashed line). 

In conclusion,  we have obtained the phase diagram  and 
characterized the crossover of inhomogeneities from strong to weak
coupling in a model of Coulomb frustrated phase separation
relevant for systems with a negative short-range compressibility. For
isotropic systems we find that the transition from the uniform phase
to the inhomogeneous phase is always first order except for a critical
point. Close to the critical point inhomogeneities are predicted to
form a BCC lattice with a subsequent transition to a triangular
lattice of rods and finally to a layered structure. The transition lines
continuously evolve into the weak coupling limit.  In the case of
anisotropic systems the transition to the inhomogeneous state 
can become second order with a tricritical point separating the second
order line at strong coupling from the first 
order line at weak coupling. Inclusion of an explicit cubic term  in
Eq.~\eqref{eq:eGLrescale}, not
considered here, will  make the phase
diagram asymmetric maintaining the topology. Thus we 
 expect our results to be qualitative valid for a wide range of
 systems with Coulomb frustrated phase separation.

\acknowledgments One of the authors (C.O.) acknowledges M. Beccaria for useful discussions.


\begin{thebibliography}{10}
\bibitem{note}
For details see Ref.~\cite{cha95} which we follow closely. We caution that
  Eq.~(\ref{eq:felatt}) differs from the one reported in Ref.~\cite{cha95}. We
  are in debt with T.C. Lubensky for confirming us that the result in the book
  has to be modified.
\end{thebibliography}

\end{document}